\documentclass[prl,amsmath,aps,twocolumn,superscriptaddress,letterpaper]{revtex4}
\usepackage{xspace,units,subfigure}
\usepackage{float}
\usepackage{amsmath}
\usepackage{amssymb}
\usepackage[normalem]{ulem}
\usepackage[pdftex]{graphicx}

\begin{document}


\title{A Test for Determining a Subdiffusive Model in Ergodic Systems from Single Trajectories}
\author{Yasmine Meroz}
\email{yasmine.meroz@weizmann.ac.il}
\affiliation{School of Chemistry, Raymond \& Beverly Sackler Faculty of 
Exact Sciences, Tel-Aviv University, Tel Aviv 69978, Israel}
\affiliation{Department of Condensed Matter Physics, Weizmann
  Institute of Science, Rehovot, 76100, Israel}
\author{Igor M. Sokolov}
\affiliation{Institut f\"{u}r Physik, Humboldt-Universit\"{a}t zu Berlin, 
Newtonstrasse 15, D-12489 Berlin, Germany }
\author{Joseph Klafter}
\affiliation{School of Chemistry, Raymond \& Beverly Sackler Faculty of 
Exact Sciences, Tel-Aviv University, Tel Aviv 69978, Israel}

\begin{abstract}
Experiments on particles' motion in living cells show that it
is often subdiffusive. This subdiffusion may be due to trapping, 
percolation-like structures, or viscoelatic behavior of the medium.  
While the models based on trapping (leading to continuous-time random walks) 
can easily be distinguished from the rest by testing their non-ergodicity, 
the latter two cases are harder to distinguish. We propose a statistical test 
for distinguishing between these two based on the space-filling
properties of trajectories, and prove its feasibility and specificity
using synthetic data. We moreover present a flow-chart for making a decision 
on a type of subdiffusion for a broader class of models.  
\end{abstract}

\maketitle
Experiments on particles' motion in living cells
aimed on understanding molecular crowding~\cite{Weiss2004,
  Caspi2000, Saxton2007, Golding2006} have unveiled
that diffusion in such environments is often anomalous, i.e. the 
mean squared displacement (MSD) does not grow
proportionally to time, $\langle x^2(t) \rangle \propto t$ , but
follows a slower pattern 
\begin{equation}
\langle  x^2(t) \rangle \propto t^\alpha
\label{eq:Anodiff}
\end{equation}
with $0< \alpha < 1 $ (subdiffusion),
and the nature of this anomaly has to be understood. Anomalous
diffusion is not only apparent in biological systems, but is found in
complex systems ranging from amorphous semiconductors~\cite{Scher1975}, goeology
~\cite{Weiss2007}, to turbulent systems~\cite{Silvestri2009}.

There are several mathematical models leading to subdiffusion, corresponding to different
physical assumptions about the structure and energy landscape of the system in
which the subdiffusive motion takes place. Since one is mostly interested
in the actual microscopic structure of the system, an important task is
working out tests which allow for distinguishing between different
models giving the same prediction for the MSD. 
The three most popular models which might be pertinent to explaining
subdiffusion in cells are: 

(i) continuous time random walk (CTRW), a mathematical model which
is physically realized in systems with traps, i.e. binding sites of
different energetic depths, a case pertinent to energetic disorder, 

(ii) diffusion on fractal structures, as exemplified by percolation, a
situation pertinent to structural disorder, and

(iii) fractional Brownian motion~\cite{Mandelbrot1968} (fBm), a
Gaussian process with stationary increments which satisfies the
following statistical properties: the process is symmetric, i.e $\langle
x_{H}(t) \rangle = 0$ where $x_{H}(0)=0$, and the MSD scales as $\langle
x ^2_{H}(t) \rangle \sim t^{2H}$ where $H$ is the Hurst exponent. Note
that $H<1/2$ leads to subdiffusion, while $H=1/2$ recovers Brownian motion.
fBm physically corresponding to
systems with predominating slow modes of motion and is realized in
viscoelastic media as exemplified by polymers and polymer networks,
where disorder does not play a leading role. 

Lastly, one has to discuss

(iv) the time-dependent diffusion coefficient (TDDC) model - normal
diffusion with a time-dependent diffusion coefficient, 
which is used to fit experimental results from, for example, FRAP (fluorescence recovery after photobleaching)
experiments~\cite{Saxton2001}. This model corresponds e.g. to a situation when the step rate is explicitly
time-dependent, and does not have a clear
physical interpretation in application to crowded media.  

The non-stationary (and non-ergodic) models of anomalous diffusion
like CTRW or TDDC 
are easily distinguished from the ergodic and
stationary models of diffusion (as exemplified by fBm or diffusion on
percolation structures) by applying tests aimed onto checking
stationarity of increments or ergodicity. At present, two of them can be
recommended: the $p$-variation test~\cite{Magdziarz2009} which can be
considered as a generalized test of temporal homogeneity of the
process, and the moving average vs. ensemble average test~\cite{Lubelski2008} 
which is a clear test for ergodicity, see \cite{Weiss2009,Kepten2011} for their
practical application. 

It is much harder to distinguish within the class of ergodic
non-Markovian processes,
i.e. to tell whether the observed subdiffusion is due to 
geometrical restrictions (e.g. percolation) or to
a viscoelastic medium (fBm). More detailed information on these two
models, and how to simulate them, appears in the Supplementary Material.
Note that both models correspond to antipersistent random walks (RWs), and may have
the same step-step (or velocity-velocity) correlation function. The
corresponding correlation function for percolation is calculated in \cite{Jacobs1990}
and on the coarse-grained level it is connected with the spectral
dimension of the percolation structure. The step-step correlation
functions are shown in the Supplementary Material.
For any physical model resulting in fBm the correlation function
follows from the spectral properties of slowest modes. Distinguishing between the
models is particularly challenging in single-molecule tracking experiments
where only one or few trajectories of motion are recorded~\cite{Ernst2012}.

One fundamental difference between the two is the probability
distribution function (PDF) of displacements which is Gaussian for fBm
but non-Gaussian for percolation, meaning that a Gaussianity test 
(i.e. in the exact relation between the second and the
higher even central moments of the PDF) may in
principle solve the problem~\cite{Tejedor2010}. 
However, the limited amount of available information is not enough to
produce a distinguishable PDF
(an example is shown in the
Supplementary Material), and moreover no analytical form is
known for the percolation PDF. Moreover, Gaussianity on its own does not shed light on the nature of
the type of constraint governing the tracer's motion, i.e whether its
motion is confined to an inhomogeneous geometrical structure, which
does not
considerably change on the time scale of the experiment, or such a
structure is absent, and the restrictions to the motion change
with time (like in the Rouse model of polymers or in single-file
diffusion). This information can be delivered by the tests of spatial
homogeneity of the corresponding motion. 
The aim of the present work is to give such a test
on a single trajectory level, and to prove its feasibility and
specificity using synthetic data for percolation and for
fBm with exactly the same MSD behavior.

Our present discussion is confined to a two-dimensional (2d) situation,
such as the diffusion of membrane proteins in the cell membrane, which 
constitutes one of the most interesting cases 
where single molecule tracking methods are used, see e.g. \cite{Skaug2011} and
references therein. Our discussion
can easily be generalized to 3d, if the data for all
three coordinates are available. Caution is advised if the data 
available corresponds to the 2d projection of a 3d
trajectory, like in \cite{Bronstein2009}, in which case our method may not be appropriate.

On the level of the RW description, the
processes with non-stationary and with stationary increments differ in how 
the clock time $t$ is translated into the steps of the problem. 
In both CTRW and TDDC the steps follow inhomogeneously in time, and
the mean number of steps $n$ taken up to time $t$ grows as $\langle n \rangle
\propto t^\alpha$, while the MSD as a function of the numer of steps
grows as $\langle  x^2(t) \rangle \propto n$. 
Thus, CTRW and TDDC models correspond to normal diffusion
if the clock time is translated into steps of the RW process.
This transformation can either follow a random process (CTRW) 
or be deterministic (TDDC). 
These processes fill space homogenously
like in normal diffusion. 
On the other hand, for fBm and for a RW on a percolation cluster,
being time-homogeneous processes (with stationary increments), 
the number of steps is always proportional to time. 
Here the fractal dimension of the trajectory
is connected to the exponent $\alpha$ characterizing anomalous diffusion,
\begin{equation}\label{eq:msd_d_w}
\langle {\bf r}^2(t) \rangle \propto t^\alpha = t^{2/d_w} \propto n^{2/d_w},
\end{equation}
where $d_w = 2/\alpha$ is called the {\it walk dimension}.
For fBm the exponent $\alpha$ is related to the Hurst
exponent by $\alpha = 2/d_w = 2H$.
The fractal dimension $d_f$ is defined through how the amount of
available sites within a radius $r$ scales with $r$: $M_n \simeq r^{d_f}$. 

Let us consider the number of sites within a radius $r$ as a
function of the average time needed to reach such a radius. We do so
by substituting the square root of the MSD's time dependance in place
of $r$: $M_n \simeq r^{d_f} \sim (n^{1/d_w})^{d_f} = n^{d_s/2}$. 

Note that we are dealing with a {\it recurrent walk}, where $M_n$
grows slower than $n$, i.e $d_f < d_w$.
In this case each of
the sites within the reachable distance is visited at least once, and
the total number of distinct visited sites behaves as $S_n \approx
M_n$, i.e~\cite{Dasgupta1994, KlafterSokolov}:
\begin{equation}
S_n \sim n^{d_f/d_w}. 
\end{equation} 
In the case of fBm the geometry is Euclidean, meaning that $d_f = 2$ or 3 in 2d or
3d respectively. In percolation, on the other hand, $d_f \approx 1.8958$ and $d_f
\approx 2.52$ when embedded in 2d and 3d respectively.  The walk dimension associated with a
RW on a 2d percolation cluster is $d_w = 2.87$.

In our approach to the problem
we propose to exploit the fact that fBm explores an Euclidean
structure ($d_f=d$, with $d$ being the dimension of space), 
while a RW on a percolation cluster explores a
fractal one with $d_f < d$.

To formulate the null-hypothesis as to how the RWer fills space, let us look at the ratio of the
average number of distinct visited sites within $n$ time steps,
$S_n$, and the space enclosed in the radius $r(n)$ which the RWer
reaches on average within the same number of time steps, $r^d(n) =
\langle {\bf r}^2 (n) \rangle ^{d/2}$. 
We examine $S_n/\langle {\bf r}^2(n) \rangle ^{d/2} \sim n^{\frac{d_f -
    d}{d_w}}$, so that our test is based on the calculation of the exponent:
\begin{equation}
\delta \equiv \frac{d_f - d}{d_w},
\end{equation}
i.e. the difference between $d_f$ and $d$ for a given $d_w$. 
We note that if the RWer fills space
homogeneously, the two quantities grow at the same rate, meaning that
the curve is expected to be flat, or $\delta=0$. Indeed for fBm $d_f = d$,
meaning that $\delta = 0$, as opposed to the case of a RW on a
percolation cluster where $d_f < d$, leading to $\delta<0$.

\begin{figure}[]
\begin{centering}
\includegraphics[width=0.45\textwidth,]{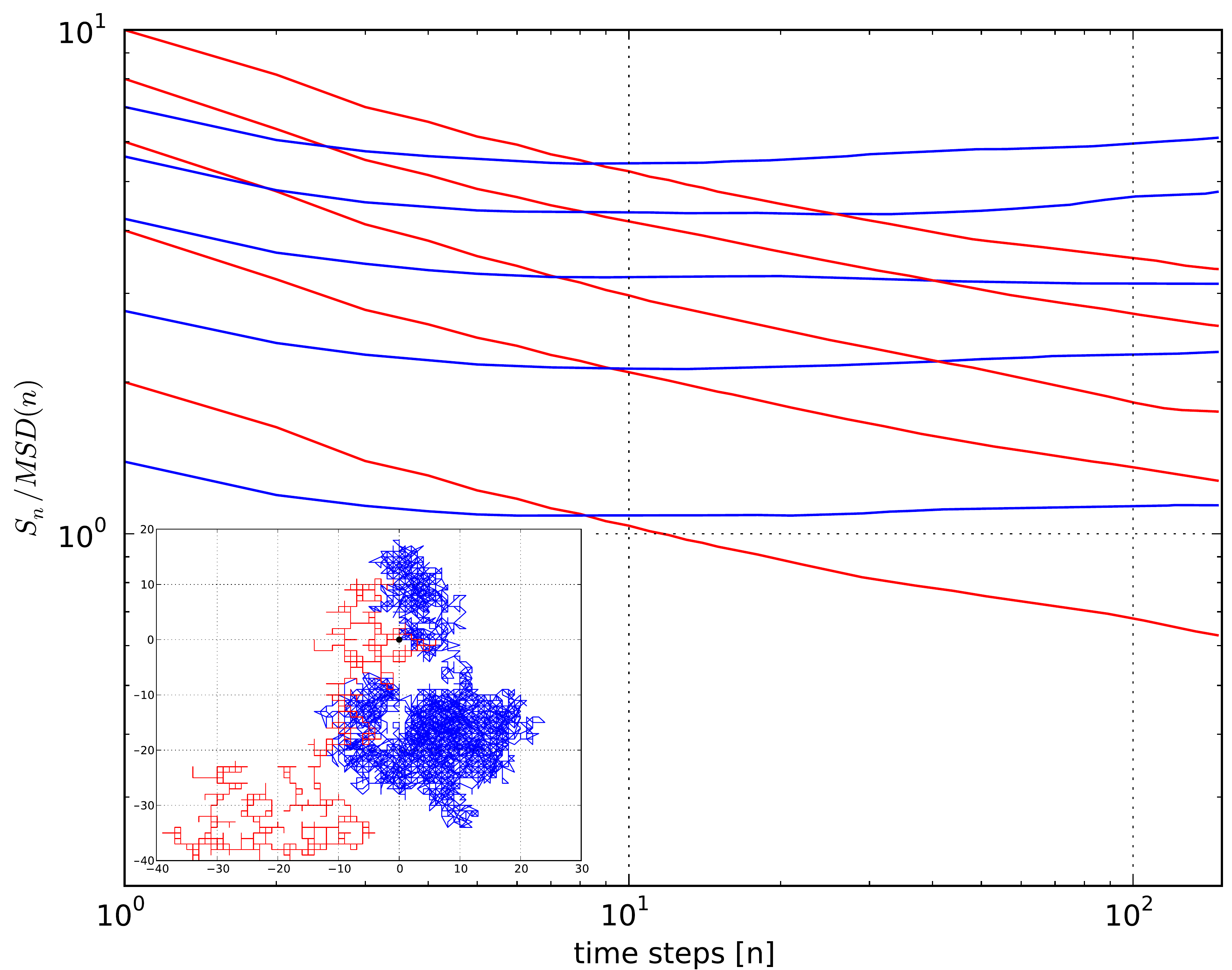}
\par\end{centering}
\caption{(color online) Participation function divided by the MSD
 ($S_n/\langle {\bf r}^2(n) \rangle ^{d/2}$ for 2d), temporally averaged with a moving window of $0 <
\tau< 150$ for 5 trajectories of fBm (blue, which flatten out) and 5 trajectories
created of a RW on a percolation cluster (red, with a clear negative
slope). All trajectories are 40000
time steps long. Two sample trajectories prior to coarse graining are shown in the inset.
\label{fig:avg_participation}}
\end{figure}

To assess the success of our test, we simulated 2d single trajectories of
fBm and of a RW on a percolation cluster at criticality. We chose the
Hurst exponent $H$ so that the MSD of the two is identical; the value 
$d_w = 2.87$ for a RW on a percolation cluster corresponds to $H=0.348$.
We modeled an experiment with optical limitations, using
a coarse grained lattice with a characteristic grain
size $\lambda$.
Two sample trajectories are
shown in the inset of Fig.~\ref{fig:avg_participation}. Note that the
trajectory of the RW on a percolation cluster is restricted here to
only horizontal and vertical directions since the percolation cluster
is based on a square lattice (see Supplementary Material for an
example of such a cluter). This is also the reason for the small
oscillations found in the ACF, also shown in the Supplementary Material.
A fast and precise generator for fractional Gaussian noise in the
antipersistent case is described in~\cite{Lowen1999}. 

Thus, our test is based on calculating
$\delta$ from the slope of $S(t)/\langle {\bf r}^2(t) \rangle^{d/2} $ on the double logarithmic scale 
and testing whether this $\delta$ is different from zero. The null-hypothesis
$\delta = 0$ corresponds to fBm, and its rejection witnesses in favor of the
percolation model. 
For the specific case of a RW on a 2d percolation cluster, we expect
$\delta = -0.037$.
Note that it is typically hard to 
detect the differences in the exponents of such magnitude on the basis of
relatively short runs. However, as will be seen in what follows,
we are in luck.

We found that for single trajectories {\it as is} the method is not sensitive 
enough due to strong noise. This noise can be reduced by looking at a moving-window
time average. Fig.~\ref{fig:avg_participation} displays
temporally averaged $S(t)/\langle
{\bf r}^2(t)\rangle^{d/2}$ in 2d, for trajectories
of a RW on a percolation cluster, and of fBm. The fBm curves flatten after the
first couple of steps, as expected, whilst the percolation ones clearly have a
negative slope.

\begin{figure}[]
\begin{centering}
\includegraphics[width=0.5\textwidth,]{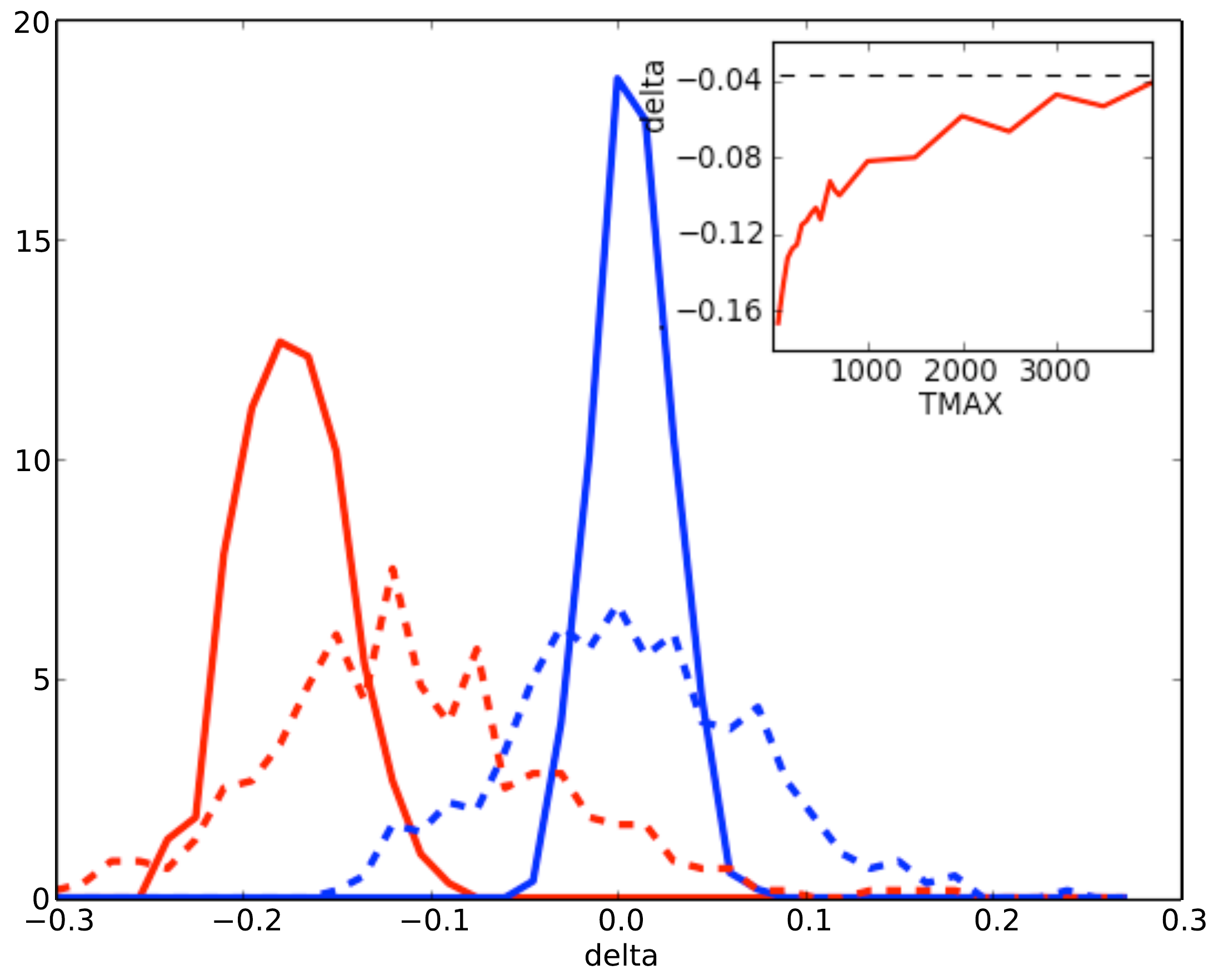}
\par\end{centering}
\caption{(color online) Distribution of $\delta$ for 400 fBm
 trajectories (blue peaks on the right) and 400 trajectories of RW on a percolation
 cluster (red peaks on the left). The trajectories are temporally
 averaged within time windows of $T_{max}=50$ time steps (solid line) and
 of $T_{max}=550$ (dashed line).
 While the fBm distributions stay centered at the expected value of
 $0$, the ones for percolation start far off (at $\approx
 -0.18$ for $T_{max}=50$) and slowly converge to the expected
 value of $-0.037$ as shown in the inset. The dashed line indicates the
expected asymptotic value. }
\label{fig:slopes_participation}
\end{figure}

\begin{figure}[]
\begin{centering}
\includegraphics[width=0.42\textwidth,]{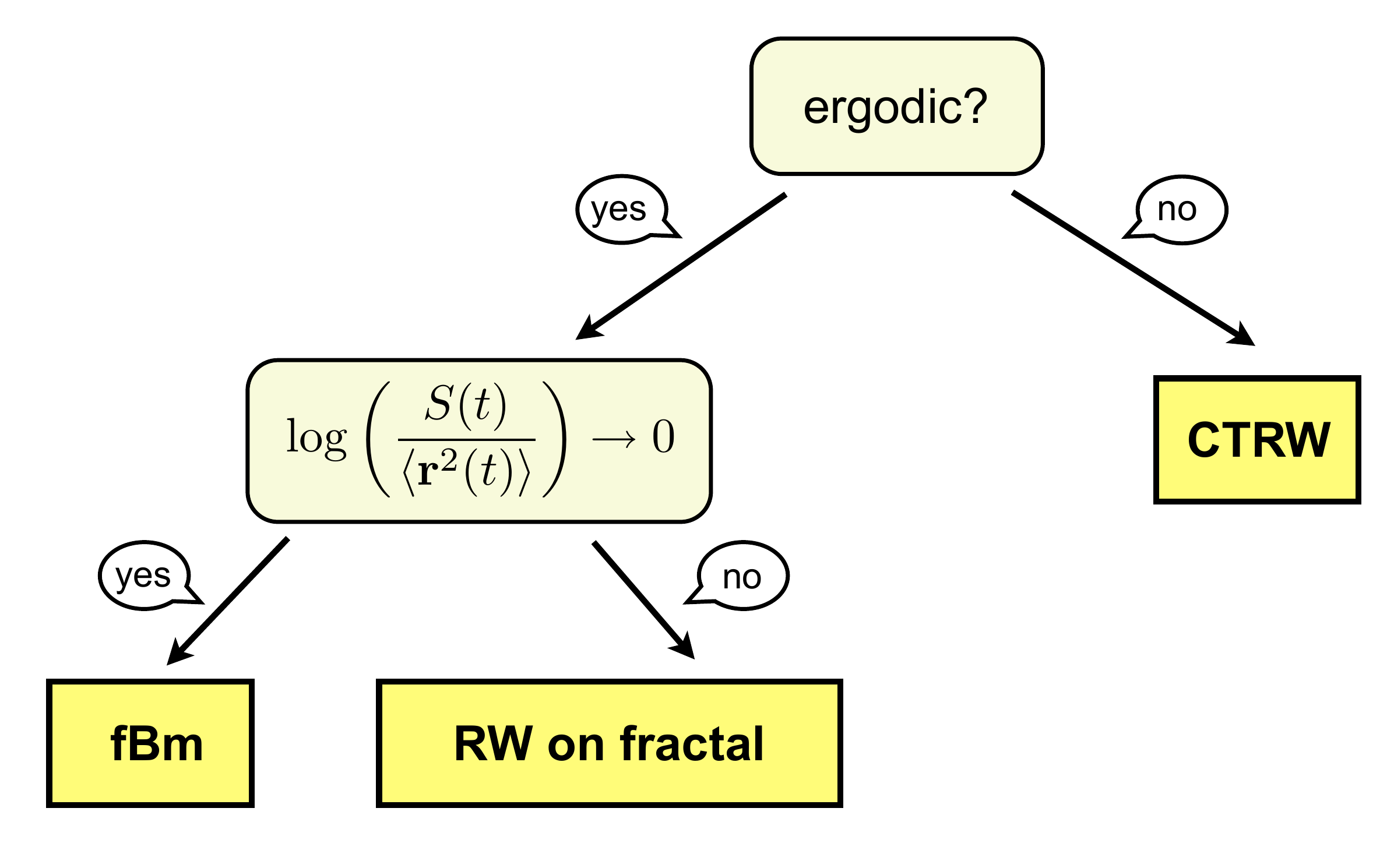}
\par\end{centering}
\caption{(color online) Flow chart of the decision process in discerning between the
 three main subdiffusive models: CTRW, fBm and a RW on a fractal
 structure. One starts by assessing the ergodicity or temporal homogeneity  
of the process. If the process if found to be non-ergodic,
 it is CTRW. If the process is ergodic, one is left to
 discriminate between fBm and a RW on a fractal structure. Here one
 analyses $S(t)/ \langle \mathbf{r}^2(t)\rangle^{d/2}$ for 2d. If this ratio is a constant, the
 process is fBm, if it decays, the process is a RW on a
 fractal structure.
\label{fig:flow_chart1}}
\end{figure}

We now fit each of these curves to a power-law and extract the
exponent corresponding to $\delta$. Fig.~\ref{fig:slopes_participation} shows the distribution
of $\delta$ resulting from $400$ fBm and percolation trajectories. 
A closer look at Fig.~\ref{fig:slopes_participation} reveals that
whilst the peak of the fBm $\delta$ distribution is centered around
$0$, the percolation distribution is not centered around $-0.037$, but
at a much larger negative number: i.e. around $-0.18$ for an averaging time
window $T_{max}=50$ and around $-0.12$ for $T_{max}=550$. This is due to 
large corrections to scaling for the percolation case (see e.g. 
\cite{Meroz2011}), which luckily play in our favour: 
for smaller $T_{max}$ the distributions are clearly
distinguishable, with no overlap, meaning also that a relatively short
trajectory is enough.
So in practice, given a single trajectory one may calculate $\delta$ for
different $T_{max}$, and see whether these are negative and
converge to a smaller negative number (pointing at a RW on a fractal), or 
tend to zero (fBm). 
We can now add this test to the test of ergodicity, building a \textit{toolbox}
to help identify the underlying physics of a given experimental
trajectory. We summarize our \textit{toolbox} in the form of a decision tree in
Fig.~\ref{fig:flow_chart1}.

One may take this toolbox one step further and consider a more
general and realistic scenario, of subordination of any two of these models, as
previously considered~\cite{Meroz2010,Weigel2011}. In a biological
cell, for example, there is no reason why one may not encounter both
energy traps (modeled with CTRW) {\it and} crowding (modeled as a RW
on a percolation cluster), i.e the problem would be modeled as a RW on
a percolation cluster, with the subordination of waiting times at each
step. This generalization is out of the scope of this paper, and will
be set forth elsewhere.

We proposed here a toolbox of tests that may be run on single
trajectories in the aim of discerning between possible physical
realities including combinations of energy traps, 
structural disorder or crowding,
and a viscoelastic medium
. We note that
not all tests may be feasible according to
the experimental setup and the type of data at hand, but it nontheless
illuminates the possibilities and gives a broader understanding.


\begin{thebibliography}{19}
\expandafter\ifx\csname natexlab\endcsname\relax\def\natexlab#1{#1}\fi
\expandafter\ifx\csname bibnamefont\endcsname\relax
  \def\bibnamefont#1{#1}\fi
\expandafter\ifx\csname bibfnamefont\endcsname\relax
  \def\bibfnamefont#1{#1}\fi
\expandafter\ifx\csname citenamefont\endcsname\relax
  \def\citenamefont#1{#1}\fi
\expandafter\ifx\csname url\endcsname\relax
  \def\url#1{\texttt{#1}}\fi
\expandafter\ifx\csname urlprefix\endcsname\relax\def\urlprefix{URL }\fi
\providecommand{\bibinfo}[2]{#2}
\providecommand{\eprint}[2][]{\url{#2}}

\bibitem[{\citenamefont{Weiss et~al.}(2004)\citenamefont{Weiss, Elsner,
  Kartberg, and Nilsson}}]{Weiss2004}
\bibinfo{author}{\bibfnamefont{M.}~\bibnamefont{Weiss}},
  \bibinfo{author}{\bibfnamefont{M.}~\bibnamefont{Elsner}},
  \bibinfo{author}{\bibfnamefont{F.}~\bibnamefont{Kartberg}}, \bibnamefont{and}
  \bibinfo{author}{\bibfnamefont{T.}~\bibnamefont{Nilsson}},
  \bibinfo{journal}{Biophys. J.} \textbf{\bibinfo{volume}{87}},
  \bibinfo{pages}{3518} (\bibinfo{year}{2004}).

\bibitem[{\citenamefont{Caspi et~al.}(2000)\citenamefont{Caspi, Granek, and
  Elbaum}}]{Caspi2000}
\bibinfo{author}{\bibfnamefont{A.}~\bibnamefont{Caspi}},
  \bibinfo{author}{\bibfnamefont{R.}~\bibnamefont{Granek}}, \bibnamefont{and}
  \bibinfo{author}{\bibfnamefont{M.}~\bibnamefont{Elbaum}},
  \bibinfo{journal}{Phys. Rev. Lett.} \textbf{\bibinfo{volume}{85}},
  \bibinfo{pages}{5655} (\bibinfo{year}{2000}).

\bibitem[{\citenamefont{Saxton}(2007)}]{Saxton2007}
\bibinfo{author}{\bibfnamefont{M.}~\bibnamefont{Saxton}},
  \bibinfo{journal}{Biophys. J.} \textbf{\bibinfo{volume}{92}},
  \bibinfo{pages}{1178} (\bibinfo{year}{2007}).

\bibitem[{\citenamefont{Golding and Cox}(2006)}]{Golding2006}
\bibinfo{author}{\bibfnamefont{I.}~\bibnamefont{Golding}} \bibnamefont{and}
  \bibinfo{author}{\bibfnamefont{E.C.}~\bibnamefont{Cox}},
  \bibinfo{journal}{Phys. Rev. Lett.} \textbf{\bibinfo{volume}{96}},
  \bibinfo{pages}{098102} (\bibinfo{year}{2006}).


\bibitem[{\citenamefont{Scher and Montroll}(2006)}]{Scher1975}
\bibinfo{author}{\bibfnamefont{H.}~\bibnamefont{Scher}} \bibnamefont{and}
  \bibinfo{author}{\bibfnamefont{E.W.}~\bibnamefont{Montroll}},
  \bibinfo{journal}{Phys. Rev. B.} \textbf{\bibinfo{volume}{12}},
  \bibinfo{pages}{2455} (\bibinfo{year}{1975}).



\bibitem[{\citenamefont{Weiss and Everett}(2007)}]{Weiss2007}
\bibinfo{author}{\bibfnamefont{C.J.}~\bibnamefont{Weiss}} \bibnamefont{and}
  \bibinfo{author}{\bibfnamefont{M.E.}~\bibnamefont{Everett}},
  \bibinfo{journal}{Journal of Geophysical Research} \textbf{\bibinfo{volume}{112}},
  \bibinfo{pages}{B08102} (\bibinfo{year}{2007}).



\bibitem[{\citenamefont{Silvestri et~al.}(2009)\citenamefont{Silvestri, Fronzoni, and Allegrini}}]{Silvestri2009}
\bibinfo{author}{\bibfnamefont{L.}~\bibnamefont{Silvestri}},
  \bibinfo{author}{\bibfnamefont{L.}~\bibnamefont{Fronzoni}},
  \bibnamefont{and} \bibinfo{author}{\bibfnamefont{P.}~\bibnamefont{Allegrini}},
  \bibinfo{journal}{Phys. Rev. Lett} \textbf{\bibinfo{volume}{102}},
  \bibinfo{pages}{014502} (\bibinfo{year}{2009}).



\bibitem[{\citenamefont{Mandelbrot
      et~al.}(1968)\citenamefont{Mandelbrot, and van Ness}}]{Mandelbrot1968}
\bibinfo{author}{\bibfnamefont{B.B.}~\bibnamefont{Mandelbrot}},\bibnamefont{and}
  \bibinfo{author}{\bibfnamefont{J.W.}~\bibnamefont{van Ness}},
  \bibinfo{journal}{SIAM Review} \textbf{\bibinfo{volume}{10}},
  \bibinfo{pages}{422} (\bibinfo{year}{1968}).

\bibitem[{\citenamefont{Saxton}(2001)}]{Saxton2001}
\bibinfo{author}{\bibfnamefont{M.}~\bibnamefont{Saxton}},
  \bibinfo{journal}{Biophys. J.} \textbf{\bibinfo{volume}{81}},
  \bibinfo{pages}{2226} (\bibinfo{year}{2001}).

\bibitem[{\citenamefont{Magdziarz et~al.}(2009)\citenamefont{Magdziarz, Weron,
  Burnecki, and Klafter}}]{Magdziarz2009}
\bibinfo{author}{\bibfnamefont{M.}~\bibnamefont{Magdziarz}},
  \bibinfo{author}{\bibfnamefont{A.}~\bibnamefont{Weron}},
  \bibinfo{author}{\bibfnamefont{K.}~\bibnamefont{Burnecki}},
  \bibnamefont{and} \bibinfo{author}{\bibfnamefont{J.}~\bibnamefont{Klafter}},
  \bibinfo{journal}{Phys. Rev. Lett} \textbf{\bibinfo{volume}{103}},
  \bibinfo{pages}{180602} (\bibinfo{year}{2009}).

\bibitem[{\citenamefont{Lubelski et~al.}(2008)\citenamefont{Lubelski, Sokolov,
  and Klafter}}]{Lubelski2008}
\bibinfo{author}{\bibfnamefont{A.}~\bibnamefont{Lubelski}},
  \bibinfo{author}{\bibfnamefont{I.~M.} \bibnamefont{Sokolov}},
  \bibnamefont{and} \bibinfo{author}{\bibfnamefont{J.}~\bibnamefont{Klafter}},
  \bibinfo{journal}{Phys. Rev. Lett} \textbf{\bibinfo{volume}{100}},
  \bibinfo{pages}{250602} (\bibinfo{year}{2008}).

\bibitem[{\citenamefont{Szymanski and Weiss}(2009)}]{Weiss2009}
\bibinfo{author}{\bibfnamefont{J.}~\bibnamefont{Szymanski}} \bibnamefont{and}
  \bibinfo{author}{\bibfnamefont{M.}~\bibnamefont{Weiss}},
  \bibinfo{journal}{Phys. Rev. Lett.} \textbf{\bibinfo{volume}{103}},
  \bibinfo{pages}{038102} (\bibinfo{year}{2009}).

\bibitem[{\citenamefont{Kepten et~al.}(2011)\citenamefont{Kepten, Bronshtein,
  and Garini}}]{Kepten2011}
\bibinfo{author}{\bibfnamefont{E.}~\bibnamefont{Kepten}},
  \bibinfo{author}{\bibfnamefont{I.}~\bibnamefont{Bronshtein}},
  \bibnamefont{and} \bibinfo{author}{\bibfnamefont{Y.}~\bibnamefont{Garini}},
  \bibinfo{journal}{Phys. Rev. E} \textbf{\bibinfo{volume}{83}},
  \bibinfo{pages}{041919} (\bibinfo{year}{2011}).

\bibitem[{\citenamefont{Jacobs and Nakanishi}(1990)}]{Jacobs1990}
\bibinfo{author}{\bibfnamefont{D.}~\bibnamefont{Jacobs}} \bibnamefont{and}
  \bibinfo{author}{\bibfnamefont{H.}~\bibnamefont{Nakanishi}},
  \bibinfo{journal}{Phys. Rev. A} \textbf{\bibinfo{volume}{41}},
  \bibinfo{pages}{706} (\bibinfo{year}{1990}).


\bibitem[{\citenamefont{Ernst et~al.}(2012)\citenamefont{Ernst, Hellmann,
  K\"{o}hler, and Weiss}}]{Ernst2012}
\bibinfo{author}{\bibfnamefont{D.}~\bibnamefont{Ernst}},
  \bibinfo{author}{\bibfnamefont{M.}~\bibnamefont{Hellmann}},
  \bibinfo{author}{\bibfnamefont{J.}~\bibnamefont{K\"{o}hler}},, \bibnamefont{and}
  \bibinfo{author}{\bibfnamefont{M.}~\bibnamefont{Weiss}},
  \bibinfo{journal}{Soft Matter} \textbf{\bibinfo{volume}{8}},
  \bibinfo{pages}{4886} (\bibinfo{year}{2012}).



\bibitem[{\citenamefont{Tejedor et~al.}(2010)\citenamefont{Tejedor, B\'{e}nichou,
  Voituriez, Jungmann, Simmel, Selhuber-Unkel, Oddershede, and Metzler}}]{Tejedor2010}
\bibinfo{author}{\bibfnamefont{V.}~\bibnamefont{Tejedor}},
  \bibinfo{author}{\bibfnamefont{O.}~\bibnamefont{B\'{e}nichou}},
  \bibinfo{author}{\bibfnamefont{R.}~\bibnamefont{Voituriez}},
  \bibinfo{author}{\bibfnamefont{R.}~\bibnamefont{Jungmann}},
  \bibinfo{author}{\bibfnamefont{F.}~\bibnamefont{Simmel}},
  \bibinfo{author}{\bibfnamefont{C.}~\bibnamefont{Selhuber-Unkel}},
  \bibinfo{author}{\bibfnamefont{L.B.}~\bibnamefont{Oddershede}}, \bibnamefont{and}
  \bibinfo{author}{\bibfnamefont{R.}~\bibnamefont{Metzler}},
  \bibinfo{journal}{Biophys. J.} \textbf{\bibinfo{volume}{98}},
  \bibinfo{pages}{1364} (\bibinfo{year}{2010}).



\bibitem[{\citenamefont{M.J. et~al.}(2011)\citenamefont{M.J., Faller, and
  Longo}}]{Skaug2011}
\bibinfo{author}{\bibfnamefont{S.}~\bibnamefont{M.J.}},
  \bibinfo{author}{\bibfnamefont{R.}~\bibnamefont{Faller}}, \bibnamefont{and}
  \bibinfo{author}{\bibfnamefont{M.}~\bibnamefont{Longo}}, \bibinfo{journal}{J.
  Chem. Phys.} \textbf{\bibinfo{volume}{134}}, \bibinfo{pages}{215101}
  (\bibinfo{year}{2011}).

\bibitem[{\citenamefont{Bronstein et~al.}(2009)\citenamefont{Bronstein, Israel,
  Kepten, Mai, Shav-Tal, Barkai, and Garini}}]{Bronstein2009}
\bibinfo{author}{\bibfnamefont{I.}~\bibnamefont{Bronstein}},
  \bibinfo{author}{\bibfnamefont{Y.}~\bibnamefont{Israel}},
  \bibinfo{author}{\bibfnamefont{E.}~\bibnamefont{Kepten}},
  \bibinfo{author}{\bibfnamefont{S.}~\bibnamefont{Mai}},
  \bibinfo{author}{\bibfnamefont{Y.}~\bibnamefont{Shav-Tal}},
  \bibinfo{author}{\bibfnamefont{E.}~\bibnamefont{Barkai}}, \bibnamefont{and}
  \bibinfo{author}{\bibfnamefont{Y.}~\bibnamefont{Garini}},
  \bibinfo{journal}{Phys. Rev. Lett.} \textbf{\bibinfo{volume}{103}},
  \bibinfo{pages}{018102} (\bibinfo{year}{2009}).

\bibitem[{\citenamefont{Haynes and Roberts}(2009)}]{Haynes2009}
\bibinfo{author}{\bibfnamefont{C.P.}~\bibnamefont{Haynes}} \bibnamefont{and}
  \bibinfo{author}{\bibfnamefont{A.~P.} \bibnamefont{Roberts}},
  \bibinfo{journal}{Phys. Rev. Lett.} \textbf{\bibinfo{volume}{103}},
  \bibinfo{pages}{020601} (\bibinfo{year}{2009}).

\bibitem[{\citenamefont{Alexander and Orbach}(1982)}]{Alexander1982}
\bibinfo{author}{\bibfnamefont{S.}~\bibnamefont{Alexander}} \bibnamefont{and}
  \bibinfo{author}{\bibfnamefont{R.}~\bibnamefont{Orbach}},
  \bibinfo{journal}{J. Phys. Lett. (Paris)} \textbf{\bibinfo{volume}{43}},
  \bibinfo{pages}{L625} (\bibinfo{year}{1982}).

\bibitem[{\citenamefont{Dasgupta et~al.}(1994)\citenamefont{Dasgupta, Ballabh,
  and Tarafdar}}]{Dasgupta1994}
\bibinfo{author}{\bibfnamefont{R.}~\bibnamefont{Dasgupta}},
  \bibinfo{author}{\bibfnamefont{T.}~\bibnamefont{Ballabh}}, \bibnamefont{and}
  \bibinfo{author}{\bibfnamefont{S.}~\bibnamefont{Tarafdar}},
  \bibinfo{journal}{Phys. Lett. A} \textbf{\bibinfo{volume}{187}},
  \bibinfo{pages}{71} (\bibinfo{year}{1994}).

\bibitem[{\citenamefont{Klafter and Sokolov}(2011)}]{KlafterSokolov}
\bibinfo{author}{\bibfnamefont{J.}~\bibnamefont{Klafter}} \bibnamefont{and}
  \bibinfo{author}{\bibfnamefont{I.M.}~\bibnamefont{Sokolov}},
  \emph{\bibinfo{title}{First Steps in Random Walks: From Tools to
  Applications}} (\bibinfo{publisher}{Oxford University Press},
  \bibinfo{year}{2011}).


\bibitem[{\citenamefont{Lowen}(1999)\citenamefont{Lowen}}]{Lowen1999}
\bibinfo{author}{\bibfnamefont{S.B.}~\bibnamefont{Lowen}},
  \bibinfo{journal}{Methodol. Comput. Appl. Probab.} \textbf{\bibinfo{volume}{1}},
  \bibinfo{pages}{445-456} (\bibinfo{year}{1999}). 



\bibitem[{\citenamefont{Meroz et~al.}(2011)\citenamefont{Meroz, Sokolov, and
  Klafter}}]{Meroz2011}
\bibinfo{author}{\bibfnamefont{Y.}~\bibnamefont{Meroz}},
  \bibinfo{author}{\bibfnamefont{I.M.}~\bibnamefont{Sokolov}}, \bibnamefont{and}
  \bibinfo{author}{\bibfnamefont{J.}~\bibnamefont{Klafter}},
  \bibinfo{journal}{Phys. Rev. E} \textbf{\bibinfo{volume}{83}},
  \bibinfo{pages}{020104} (\bibinfo{year}{2011}).

\bibitem[{\citenamefont{Meroz et~al.}(2010)\citenamefont{Meroz, Sokolov, and
  Klafter}}]{Meroz2010}
\bibinfo{author}{\bibfnamefont{Y.}~\bibnamefont{Meroz}},
  \bibinfo{author}{\bibfnamefont{I.M}~\bibnamefont{Sokolov}}, \bibnamefont{and}
  \bibinfo{author}{\bibfnamefont{J.}~\bibnamefont{Klafter}},
  \bibinfo{journal}{Phys. Rev. E} \textbf{\bibinfo{volume}{81}},
  \bibinfo{pages}{010101} (\bibinfo{year}{2010}).

\bibitem[{\citenamefont{Weigel et~al.}(2011)\citenamefont{Weigel, Simon,
  Tamkun, and Krapf}}]{Weigel2011}
\bibinfo{author}{\bibfnamefont{A.}~\bibnamefont{Weigel}},
  \bibinfo{author}{\bibfnamefont{B.}~\bibnamefont{Simon}},
  \bibinfo{author}{\bibfnamefont{M.}~\bibnamefont{Tamkun}}, \bibnamefont{and}
  \bibinfo{author}{\bibfnamefont{D.}~\bibnamefont{Krapf}},
  \bibinfo{journal}{Proc. Nat. Acad. Sci.} \textbf{\bibinfo{volume}{108}},
  \bibinfo{pages}{6438} (\bibinfo{year}{2011}).



\end{thebibliography}
\end{document}



\title{A Test for Determining a Subdiffusive Model in Ergodic Systems
  from Single Trajectories - Supplementary Material}
\author{Yasmine Meroz}
\email{yasmine.meroz@weizmann.ac.il}
\affiliation{School of Chemistry, Raymond \& Beverly Sackler Faculty of 
Exact Sciences, Tel-Aviv University, Tel Aviv 69978, Israel}
\affiliation{Department of Condensed Matter Physics, Weizmann
  Institute of Science, Rehovot, 76100, Israel}
\author{Igor M. Sokolov}
\affiliation{Institut f\"{u}r Physik, Humboldt-Universit\"{a}t zu Berlin, 
Newtonstrasse 15, D-12489 Berlin, Germany }
\author{Joseph Klafter}
\affiliation{School of Chemistry, Raymond \& Beverly Sackler Faculty of 
Exact Sciences, Tel-Aviv University, Tel Aviv 69978, Israel}

\maketitle
\section{Description of the models}
\begin{figure}[]
\begin{centering}
\includegraphics[width=0.4\textwidth,]{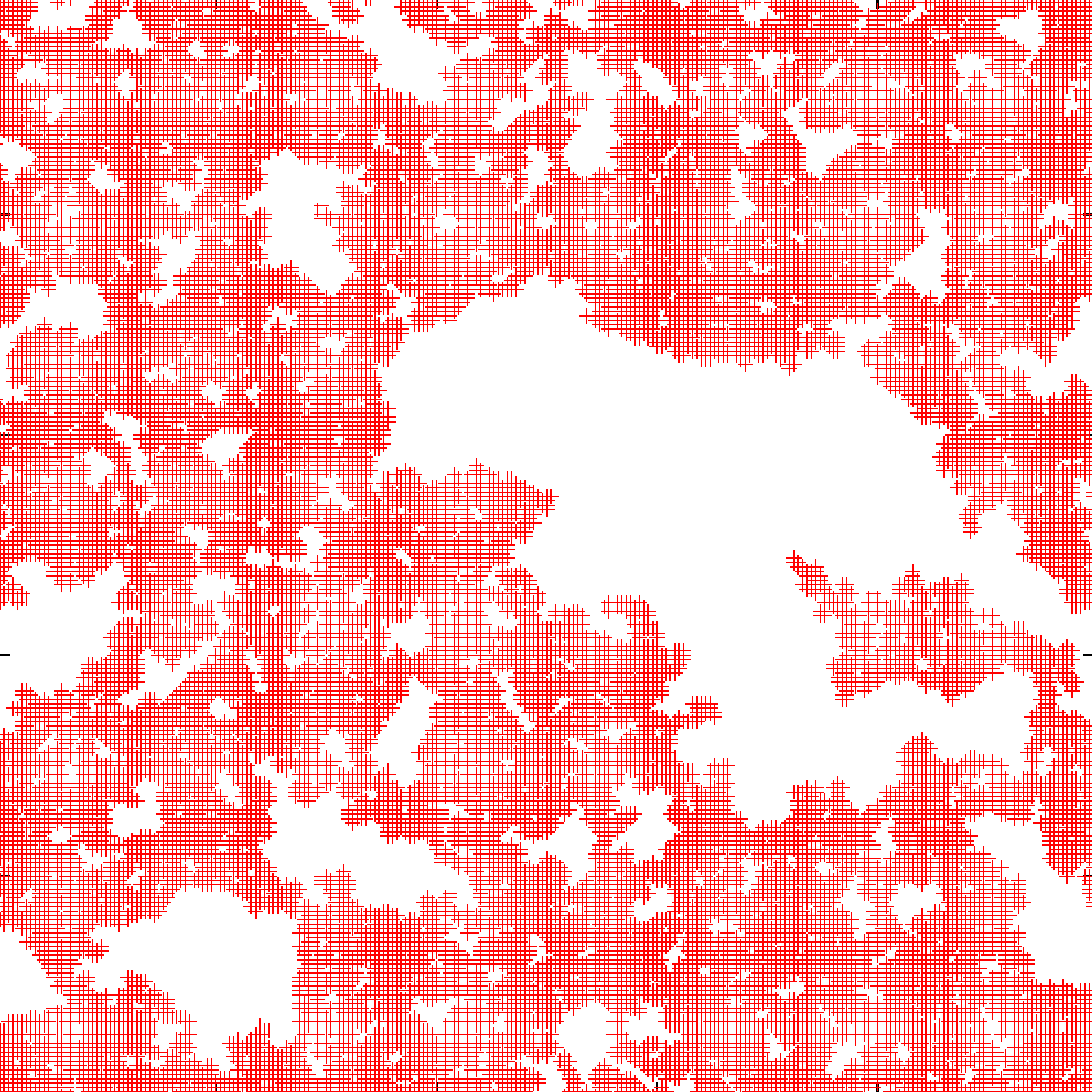}
\par\end{centering}
\caption{An example of a two-dimensional site percolcation cluster at
  criticality.
\label{fig:percolation_cluster}}
\end{figure}

Let us first describe how one simulates a percolaton cluster at
criticalitiy~\cite{KlafterSokolov}. We take a two dimensional lattice, and start removing
sites. The critical concentration of sites for a two-dimensional
square lattice is $p_c = 0.5$. For a higher concentration there are
multiple ways for a RW to reach from one side of the lattice to the
other, for a lower concentration so many sites have been removed that
the two sides of the lattice are detached. The critical concentration
is that in which, on average, removing one more site will detach the
lattice sides. A RWer is allowed to walk between the remaining sites,
with the remaining bonds.
Fig.~\ref{fig:percolation_cluster} shows and example of such a percolation
cluster at criticality. Note the empty spaces at different
(scale-free) sizes. It should also be noted that since the cluster is
based on a square lattice, the trajectory of a RW on such a cluster
will also be restricted to only horizontal and vertical steps.

FBm is generated from fractional Gaussian noise, just as Brownian
motion is generated
from white noise. Mandelbrot and van Ness~\cite{Mandelbrot1968} defined fBm as:
\begin{equation}
B_H(t) \equiv \frac{1}{\Gamma(H+\frac{1}{2})} \left (
  \int_0^t{(t-\tau)^{H-\frac{1}{2}} dB(\tau) } +
  \int_{-\infty}^{0}{[(t-\tau)^{H-\frac{1}{2}}  -
    (-\tau)^{H-\frac{1}{2}} ]  dB(\tau) }\right ), 
\end{equation}
where $\Gamma$ is the Gamma functin and $0<H<1$ is the Hurst
parameter. Note that $H=\frac{1}{2}$ recovers ordinary Brownian motion.

As for fBm, quite a few methods of simulation exist. We chose one of
the most popular methods which uses a discrete Fourier transform. The
method is thoroughly described in~\cite{Lowen1999}.

\section{PDF and ACF are inadequate to distinguish between the models}
\begin{figure}[]
\begin{centering}
\includegraphics[width=0.45\textwidth,]{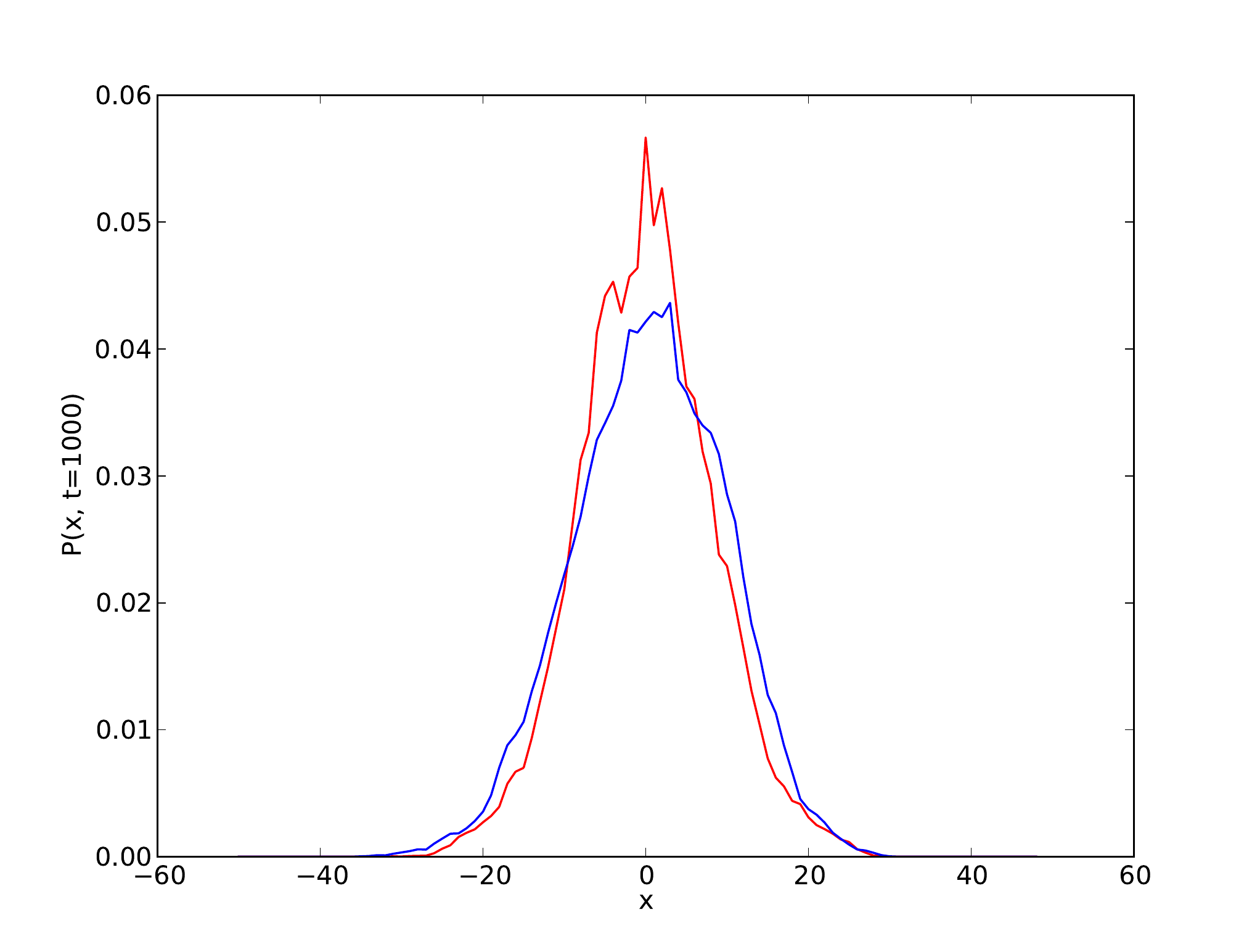}
\par\end{centering}
\caption{(color online) Probability distribution function calculated
  form a single trajectory  of fBm (blue) and of a RW on a percolation cluster (red). Both trajectories are 40000
time steps long.
\label{fig:pdf1}}
\end{figure}

\begin{figure}[]
\begin{centering}
\includegraphics[width=0.45\textwidth,]{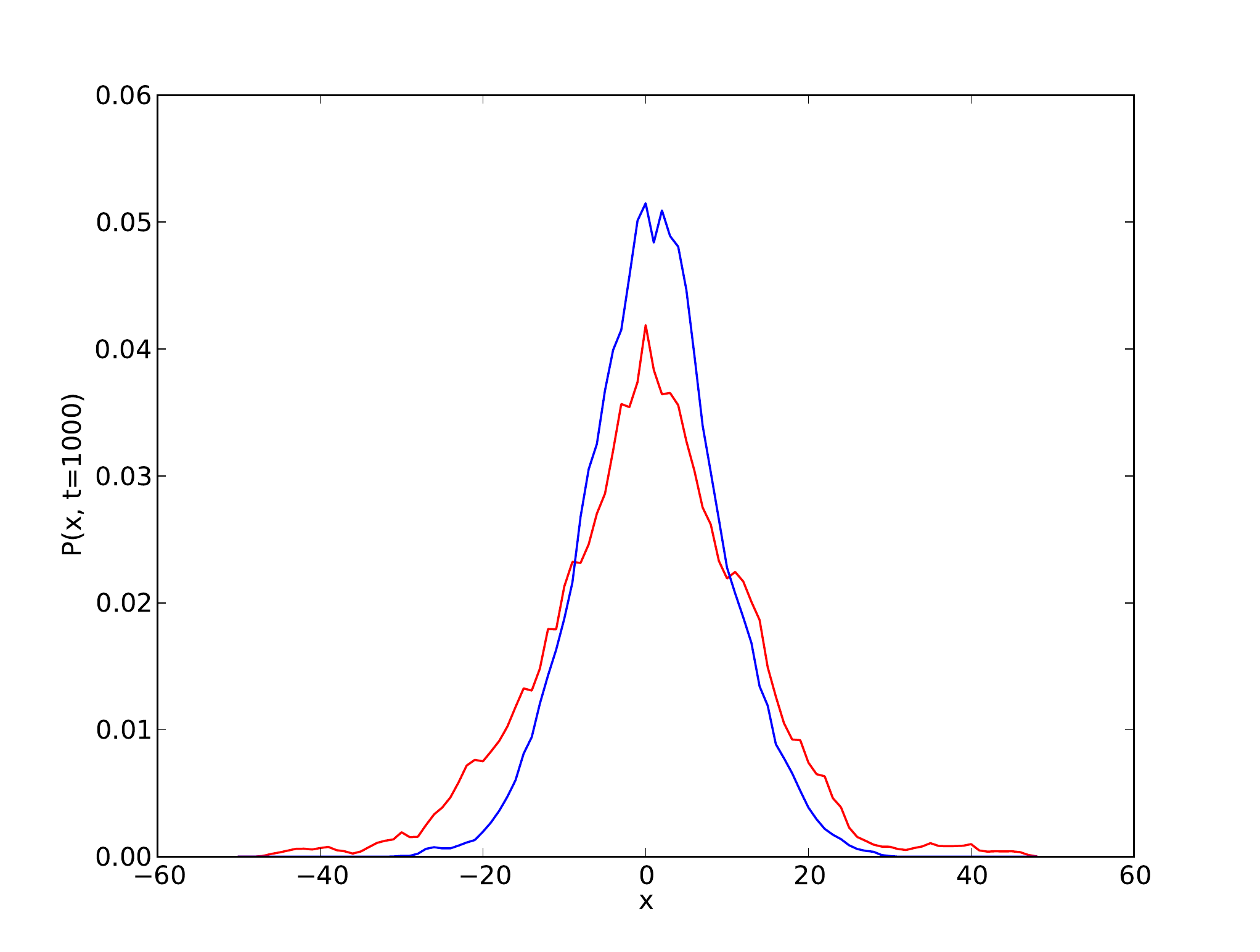}
\par\end{centering}
\caption{(color online) (color online) Probability distribution function calculated
  form a single trajectory  of fBm (blue) and of a RW on a percolation cluster (red). Both trajectories are 40000
time steps long.
\label{fig:pdf2}}
\end{figure}

\begin{figure}[]
\begin{centering}
\includegraphics[width=0.45\textwidth,]{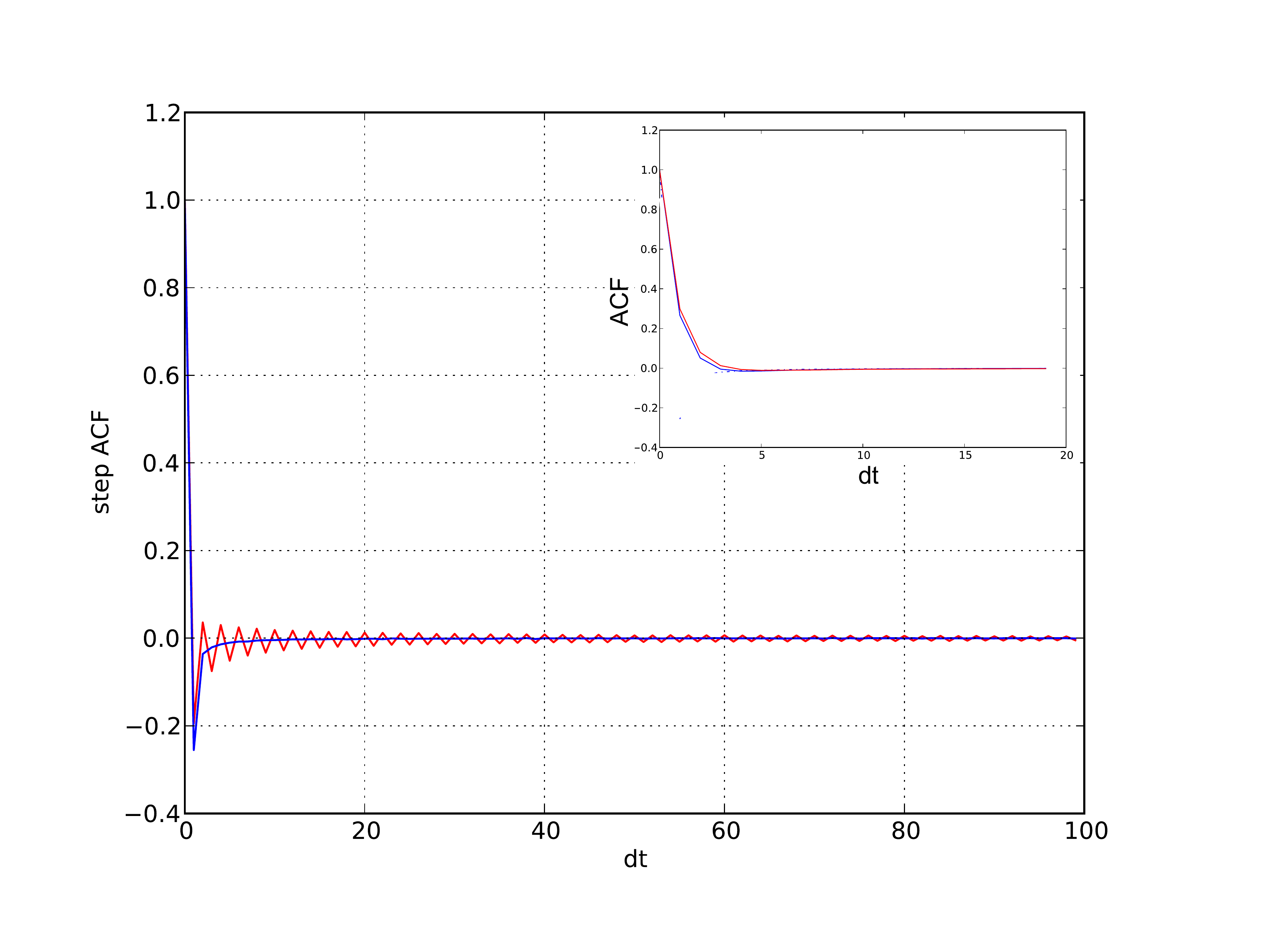}
\par\end{centering}
\caption{(color online) The step-step ACF of  fBm
  (blue) and of a RW on a percolation cluster (red). In the inset: the
  ACF of the two but after adding waiting times
  from an exponenetial distribution (a more realistic view than
  identically timed steps). The slight oscillations of the RW on a
  percolation cluster disappear, and the two ACFs are indistinguishable.}
\label{fig:ACF}
\end{figure}

Fig.~\ref{fig:pdf1} and \ref{fig:pdf2} show the PDFs for two pairs of trajectories of fBm and
for RW on a percolation cluster. The behavior of these two examples is
drastically different, and shows that obtaining the known final forms
of the distribution needs large ensembles.
Fig.~\ref{fig:ACF} shows the ACF for fBm and for a
RW on a percolation cluster. The two are identical apart from very
small oscillations apparent in the RW on a percolation cluster. Adding
exponentially distributed waiting times to the steps (a more realistic
case than equally timed steps), eliminates these oscillations,
yielding undistinguishable ACFs.

